\documentclass[letterpaper,english]{elsarticle}
\usepackage[T1]{fontenc}
\usepackage[latin9]{inputenc}
\usepackage{amsmath}
\usepackage{amssymb}
\usepackage{esint}

\makeatletter

\pdfpageheight\paperheight
\pdfpagewidth\paperwidth

\makeatother

\usepackage{babel}
\begin{document}

\begin{frontmatter}{}

\title{Mellin transforming the minimal model CFTs:\\
AdS/CFT at strong curvature}

\author{David A. Lowe}

\address{Physics Department, Brown University, Providence, RI, 02912, USA}
\begin{abstract}
Mack has conjectured that all conformal field theories are equivalent
to string theories. We explore the example of the two-dimensional
minimal model CFTs and confirm that the Mellin transformed amplitudes
have the desired properties of string theory in three-dimensional
anti-de Sitter spacetime.
\end{abstract}

\end{frontmatter}{}

\section{Introduction}

Mack \citep{Mack:2009gy,Mack:2009mi} has long advocated that Mellin
transforming CFT correlators is useful for efficiently understanding
the structure of these theories. The examples that have primarily
been studied in the past have been higher dimensional CFTs where some
kind of perturbative expansion or partial wave expansion is needed
to solve for the nontrivial CFT correlation functions.

In the present paper, we consider a two-dimensional CFT where the
nontrivial 4-point correlation functions have analytic expressions.
These expressions may be written as sums over products of chiral conformal
blocks. The Mellin transform of these conformal blocks maps naturally
into Koba-Nielsen open string amplitudes \citep{Koba:1969kh}, for
special values of the kinematic variables. This leads us to conjecture
that the string theory dual to the CFT is equivalent to an open string
description, with many features similar to the KLT construction \citep{Kawai:1985xq}.
The analogs of the Mandelstam kinematic invariants of the boundary
S-matrix of the string theory provide coordinates for Mellin space.
The analytic expressions for the 4-point functions allow us to study
the transform without making perturbative expansions, or partial wave
expansions, avoiding subtle issues of convergence.

For the minimal model CFTs in two dimensions, we find that the Mellin
representation of the conformal blocks has simple poles along a set
of Regge trajectories, with residues polynomial in the kinematic variables.
By construction the amplitude satisfies unitarity, factorization and
crossing symmetry. These are the necessary and sufficient requirements
for the theory to have a dual interpretation as a dual resonance model.
This construction should be viewed as a generalization of the Veneziano
approach \citep{Veneziano:1968yb} to constructing open string theory
from the same basic physical requirements. 

The most economical interpretation in terms of on-shell string scattering
comes from string theory in a three-dimensional AdS geometry. The
general mapping of CFT amplitudes to AdS bulk amplitudes is described
in \citep{Mack:2009mi}. Here the geometry is curved on scales of
order the string length, so there is no low-energy gravity approximation.
Nor do we have a perturbative string coupling to control a genus expansion.
Nevertheless, the polynomial structure of the residues in the Mellin
representation indicates the string representation is local in this
three-dimensional spacetime.

\section{Mellin transforming the CFT}

For a rational two-dimensional conformal field theory, the four-point
functions can be written
\begin{equation}
\left\langle \phi_{1}(0,0)\phi_{2}(z,\bar{z})\phi_{3}(1,1)\phi_{4}(\infty,\infty)\right\rangle =\sum_{i}X_{i}I_{i}(z)\bar{I_{i}(z)}\label{eq:blocks}
\end{equation}
where the sum is over the full set of primary operators, and the $I_{i}$
are conformal blocks. Here we have used global conformal symmetry
to fix the positions of three of the operators. 

For the minimal models there exist explicit integral expressions for
the $I_{i}(z)$ derived in \citep{Dotsenko:1984nm,Dotsenko:1984ad}
using the Coulomb gas representation of the conformal field theory.
These take the form of chiral correlators 
\begin{align}
I_{i}(z) & =\oint_{C_{1}}du_{1}\cdots\oint_{C_{n}}du_{n}\oint_{D_{1}}dv_{1}\cdots\oint_{D_{m}}dv_{m}\left\langle V_{\alpha_{1}}(0)V_{\alpha_{2}}(z)V_{\alpha_{3}}(1)V_{\alpha_{4}}(\infty)\prod_{i=1}^{n}J_{-}(u_{i})\prod_{j=1}^{m}J_{+}(v_{j})\right\rangle \nonumber \\
 & =N\,z^{2\alpha_{1}\alpha_{2}}(1-z)^{2\alpha_{3}\alpha_{2}}\oint_{C_{1}}du_{1}\cdots\oint_{C_{n}}du_{n}\oint_{D_{1}}dv_{1}\cdots\oint_{D_{m}}dv_{m}\prod_{i=1}^{n}u_{i}^{2\alpha_{-}\alpha_{1}}(u_{i}-1)^{2\alpha_{-}\alpha_{3}}(u_{i}-z)^{2\alpha_{-}\alpha_{2}}\nonumber \\
\times & \prod_{j=1}^{m}v_{j}^{2\alpha_{+}\alpha_{1}}(v_{j}-1)^{2\alpha_{+}\alpha_{3}}(v_{j}-z)^{2\alpha_{+}\alpha_{2}}\prod_{i<j}^{n}(u_{i}-u_{j})^{2\alpha_{-}^{2}}\prod_{i<j}^{m}(v_{i}-v_{j})^{2\alpha_{+}^{2}}\prod_{i,j}(u_{i}-v_{j})^{-2}\label{eq:chiralblock}
\end{align}
The Coulomb gas charges
\[
\alpha_{n,m}=\frac{1}{2}(1-n)\alpha_{-}+\frac{1}{2}(1-m)\alpha_{+}
\]
determine the operator conformal weights
\[
\Delta_{n,m}=\frac{1}{4}\left(\left(\alpha_{-}n-\alpha_{+}m\right)^{2}-\left(\alpha_{-}+\alpha_{+}\right)^{2}\right)
\]
where
\[
\alpha_{\pm}=\alpha_{0}\pm\sqrt{\alpha_{0}^{2}+1}
\]
and $\alpha_{0}$ is determined by the central charge
\[
c=1-24\alpha_{0}^{2}.
\]
In the above $N$ is a normalization constant. The specification of
the contours is described in detail in \citep{Dotsenko:1984nm,Dotsenko:1984ad}.
The contours may be chosen as intervals along the real axis. For a
given set of operators $V_{i}(z)$ there is a minimal choice for the
set of screening operators $J_{\pm}$ which yield non-vanishing integrals.
The number of independent contours depends on the set of $V_{i}(z)$. 

The derivation of \eqref{eq:blocks} and \eqref{eq:chiralblock} \citep{Dotsenko:1984nm,Dotsenko:1984ad}
proceeds via an identical analytic continuation to the mapping of
tree-level open string amplitudes into closed string amplitudes \citep{Kawai:1985xq}.
The mapping is guaranteed for worldsheets with disc topology, since
the process may be viewed either as creation of a closed string from
the vacuum (with other vertex operator insertions) or as an open string
being created and then subsequently annihilated. This motivates the
conjecture that the minimal model CFT can likewise be viewed as arising
from a more fundamental chiral description.

Our interest then will be to study the Mellin transform of the basic
chiral blocks of the minimal model CFTs in order to study the open
string description. One may then reinterpret the two-dimensional CFT
amplitudes as a definition of a dual holographic theory in three-dimensional
anti-de Sitter spacetime following \citep{Mack:2009mi}. The Mellin
transformed amplitudes are naturally interpreted as boundary S-matrix
elements of the bulk theory, expressed in terms of kinematic variables.
Our goal is to study the analytic properties of these Mellin amplitudes,
and show that the expected structure of a dual resonance model emerges,
in particular Regge trajectories with vanishing dispersion, and interactions
polynomial in the kinematic variables. 

To define the Mellin transform of $I_{i}(z)$ we follow Mack's suggestion
\citep{Mack:2009mi} , eqn (69), to consider the chiral transform
\begin{equation}
I_{i}(z)=\frac{1}{\left(2\pi i\right)^{2}}\int_{-i\infty+c}^{i\infty+c}d\beta_{12}\int_{-i\infty+c'}^{i\infty+c'}d\beta_{23}\,z^{-\beta_{12}}(1-z)^{-\beta_{23}}\hat{M}_{i}\left(\left\{ \beta_{ij}\right\} \right)\label{eq:mellindef}
\end{equation}
to define the reduced Mellin amplitude $\hat{M}_{i}$. Here $c$ and
$c'$ are constants chosen so the integral converges. Further details
of the interpretation of the Mellin amplitudes can be found in \citep{Mack:2009mi}.

Our first task is to invert the formula \eqref{eq:mellindef} and
solve for $\hat{M}_{i}$. Consider the the integrand in \eqref{eq:chiralblock},
\begin{align*}
J_{i}(z) & =z^{2\alpha_{1}\alpha_{2}}(1-z)^{2\alpha_{3}\alpha_{2}}\prod_{i=1}^{n}u_{i}^{2\alpha_{-}\alpha_{1}}(u_{i}-1)^{2\alpha_{-}\alpha_{3}}(u_{i}-z)^{2\alpha_{-}\alpha_{2}}\\
 & \times\prod_{j=1}^{m}v_{j}^{2\alpha_{+}\alpha_{1}}(v_{j}-1)^{2\alpha_{+}\alpha_{3}}(v_{j}-z)^{2\alpha_{+}\alpha_{2}}\prod_{i<j}^{n}(u_{i}-u_{j})^{2\alpha_{-}^{2}}\prod_{i<j}^{m}(v_{i}-v_{j})^{2\alpha_{+}^{2}}\prod_{i,j}(u_{i}-v_{j})^{-2}
\end{align*}
This is a special case of the integrand that appears in the general
Koba-Nielsen expression for the $(4+m+n)$-open string scattering
amplitudes \citep{Koba:1969kh}.

The Mellin transform of open string amplitudes has been studied by
Stieberger and Taylor \citep{Stieberger:2013hza}. In order to write
their expression in a more symmetric form, it is written as a distribution
to be integrated over an overcomplete set of cross-ratios. Let us
label the $N=(4+m+n)$ points by $z_{i}$. Defining
\[
u_{i,j}=\frac{(z_{i}-z_{j})(z_{i-1}-z_{j+1})}{(z_{i}-z_{j+1})(z_{i-1}-z_{j})}
\]
we obtain a basis for the $N(N-3)/2$ anharmonic ratios. In this formula
$(i,j)$ run over pairs conjugate to the kinematic channels, corresponding
to the range $i=2,j=3,\cdots,N-1$ and $i=3,\cdots,N-1<j=4,\cdots,N$.
The notation $P$ is denotes this set of channels which corresponds
to the set of independent kinematic invariants of the string amplitude
$s_{i,j}$ which are analogs of the usual flat spacetime Mandelstam
variables $\left(k_{i}+\cdots+k_{j}\right)^{2}$. 

The cross-ratios satisfy constraint equations, leaving only $N-3$
free variables to be integrated over. The main result found in \citep{Stieberger:2013hza}
is 
\[
\prod_{(i,j)\in P}u_{i,j}^{n_{i,j}}\theta(1-u_{i,j})\delta\left(\left\{ u_{k,l}\right\} \right)=\frac{1}{\left(2\pi i\right)^{m}}\left(\prod_{(i,j)\in P}\int_{-i\infty+c}^{i\infty+c}ds_{i,j}u_{i,j}^{-s_{i,j}}\right)B_{N}\left(\left\{ s_{k,l}\right\} ,\left\{ n_{k,l}\right\} \right)
\]
Here the $n_{i,j}$ are integers subject to the constraint $n_{i,j}=n_{j+1,i-1}$
and $n_{k,N}=n_{1,k-1}$. The constraint delta function is
\[
\delta\left(\left\{ u_{i,j}\right\} \right)=\prod_{P}\,'\delta\left(u_{P}-1+\prod_{\tilde{P}}u_{\tilde{P}}\right)
\]
where the prime denotes the exclusion of the $(2,k)$ channels and
$\tilde{P}$ denotes the set of channels incompatible with $P$. Incompatible
channels cannot appear simultaneously with a given channel $P$ and
satisfy the condition that if $u_{P}=0$ then $u_{\tilde{P}}=1$.
For more details on the definition of incompatible channels see \citep{Stieberger:2013hza}
and references therein.

The $B_{N}\left(\left\{ s_{n,l}\right\} ,\left\{ n_{k,l}\right\} \right)$
are just the Koba-Nielsen open string amplitudes \citep{Koba:1969kh}
written as a Mellin transform
\[
B_{N}\left(\left\{ s_{k,l}\right\} ,\left\{ n_{k,l}\right\} \right)=\left(\prod_{i,j\in P}\int_{0}^{\infty}du_{i,j}u_{i,j}^{s_{i,j}-1+n_{i,j}}\theta(1-u_{i,j})\right)\delta\left(\left\{ u_{k,l}\right\} \right)
\]

For the case at hand, considering the transform of a conformal block
in a minimal model, only two of the $s_{i,j}$ are independent, corresponding
to the $\beta_{12}$ and $\beta_{23}$ of \eqref{eq:mellindef}. The
remaining $s_{i,j}$ are fixed by the exponents in \eqref{eq:chiralblock}
(after performing the change of variables from $z_{i}$ to $u_{i,j}$
\citep{Stieberger:2013hza}).

From the known properties of the Koba-Nielsen amplitudes, we conclude
the general minimal model Mellin transforms have infinite towers of
single poles in the independent kinematic variables. Moreover the
residues at these poles are polynomial in the other variables.

\section{Example: critical Ising model}

To provide an explicit example of the above correspondence let us
consider the conformal blocks that appear in the 4-spin correlator
in the Ising model. The conformal blocks \citep{BELAVIN1984333} are
\[
\mathcal{F}_{\pm}(x)=\frac{1}{\sqrt{2}}\left(x(1-x)\right)^{-1/8}\sqrt{1\pm\sqrt{1-x}}
\]

To define the Mellin transformed open string amplitude we consider
the integral

\[
B_{3}^{\pm}(\beta_{12},\beta_{23})=\int_{0}^{1}dxx^{\beta_{12}-1}(1-x)^{\beta_{23}-1}\mathcal{F}_{\pm}(x)
\]
This gives the expression
\[
B_{3}^{+}=\sqrt{2}\Gamma\left(\beta_{12}-\frac{1}{8}\right)\Gamma\left(2\beta_{23}-\frac{1}{4}\right)\,_{2}\tilde{F}_{1}\left(\frac{5}{8}-\beta_{12},2\beta_{23}-\frac{1}{4};\beta_{12}+2\beta_{23}-\frac{3}{8};-1\right)
\]
The analytic structure is simple. The regularized hypergeometric functions
$\,_{2}\tilde{F}_{1}$ have no poles as a function of $\beta_{12}$
and $\beta_{23}$. Single poles appear along the two Regge trajectories
(with $m,n$ positive integers)
\begin{eqnarray*}
\beta_{12} & = & \frac{9}{8}-n\\
\beta_{23} & = & \frac{5}{8}-\frac{m}{2}
\end{eqnarray*}
with polynomial residues in the other variable. 

For the other block we get
\[
B_{3}^{-}=\sqrt{2}\Gamma\left(\beta_{12}+\frac{3}{8}\right)\Gamma\left(2\beta_{23}-\frac{1}{4}\right)\,_{2}\tilde{F}_{1}\left(\frac{9}{8}-\beta_{12},2\beta_{23}-\frac{1}{4};\beta_{12}+2\beta_{23}+\frac{1}{8};-1\right)
\]
Single poles appear at
\begin{eqnarray*}
\beta_{12} & = & \frac{5}{8}-n\\
\beta_{23} & = & \frac{5}{8}-\frac{m}{2}
\end{eqnarray*}
with polynomial residues in the other variable. These results indicate
infinite towers of massive states contribute to the amplitude in the
string theory dual.

\section{Discussion}

Our main result is that minimal model correlation functions can be
viewed as a computed by a kind of open string theory with meromorphic
Mellin amplitudes. This is in accord with Mack's conjecture that all
conformal field theories are dual to string theories. In the present
paper, the string theory is to be treated at the purely classical
level, with no sum over topologies beyond the disk. This may be viewed
as a limit where the string coupling goes to zero. Alternatively,
the string worldsheet theory may be viewed as a chiral gravity theory.

One can nevertheless propose a duality between the minimal models
and some theory containing gravity in three-dimensional anti-de Sitter
spacetime. Each primary of the conformal group maps to a bulk field.
For scalar operators, the mass of the field in the bulk is related
to the conformal dimension via
\begin{equation}
\Delta=\frac{1}{2}\pm\sqrt{\frac{1}{4}+m^{2}R^{2}}\label{eq:mapping}
\end{equation}
where $R$ is the AdS radius of curvature in string units. Analogous
formulas exist for general spin. The on-shell boundary S-matrix written
in terms of kinematic variables for this bulk theory may then be identified
with the Mellin amplitude of the boundary conformal field theory.
This is described in detail in \citep{Mack:2009mi}. The infinite
sequences of poles in the chiral Mellin amplitudes considered above
leads to the prediction that the dual bulk string theory contains
infinite towers of massive string states. 

This leads to the dramatic conclusion that the bulk theory may be
viewed as a local theory. The Mellin amplitudes are meromorphic with
residues that are polynomial in the other kinematic variables, as
is the case in the familiar critical string theory amplitudes. Of
course the underlying CFT structure guarantees crossing symmetry and
factorization. But bulk locality in the sense of this analytic structure
comes as a complete surprise. In the past, locality would emerged
only in a large $N$ limit where the bulk correlators have a perturbative
expansion \citep{Hamilton:2005ju,Hamilton:2006az,Kabat:2011rz}. 

The minimal model CFTs have only a finite number of primary fields.
Each of these primary fields nevertheless has an infinite number of
descendant fields, obtained by the action of the Virasoro algebra.
Null states and their descendants may be truncated, but infinite towers
of states remain. The Mellin amplitude contains a Regge trajectory
associated with each primary, with the higher satellite poles corresponding
to the descendant states. 

It has been conjectured \citep{Castro:2011zq} that the simplest minimal
model, the critical Ising model, is dual to Einstein gravity in three-dimensional
anti-de Sitter spacetime. The correspondence there relied on a matching
of partition functions. At first sight, this seems surprising in the
present context, because the reduced Mellin amplitudes have poles
along entire Regge trajectories. Since the Einstein gravity theory
is to be studied at strong curvature and large Newton constant, the
matching of the proposal is hard to check. It might be that the Einstein
action is sufficient to describe black hole states with the properties
of the Regge trajectories described above.

The present results should be viewed as complementary to the results
of Mack \citep{Mack:2009mi,Mack:2009gy}, where an expansion in terms
of Euclidean partial waves leads to similar conclusions regarding
the analytic structure of the Mellin amplitudes. The new feature in
the present work is the fact that a conformal block sums over an infinite
number of Euclidean partial waves. Nevertheless, the analytic structure
is preserved. We also note our conjecture that the minimal models
have an open string interpretation in three-dimensional anti-de Sitter
spacetime goes somewhat beyond the original conjecture of Mack \citep{Mack:2009gy},
which the additional assumption that the boundary CFT descended from
a higher dimensional CFT.

Starting with a local bulk theory with a perturbative expansion, it
was shown in \citep{Fitzpatrick:2011ia} that the Mellin transform
provides a useful description of the bulk correlators. In the present
work there is no perturbative expansion for the bulk correlators.
In fact the entire bulk description is made largely at the level of
mappings of operators according to representations of the conformal
group. Given the AdS radius of curvature will be of order one in string
units, for minimal model duals, there will not be a low energy limit
where gravity decouples. However it seems likely a version of string
field theory will be applicable, and these results indicate the interactions
will have a local interpretation.

\subsection*{Acknowledgements}

I thank Marcus Spradlin and Steven Avery for helpful discussions.
The research of D.L. was supported in part by DOE grant DE-SC0010010.
\bibliographystyle{apsrev}
\bibliography{ising}

\end{document}